\documentclass[aps,11pt]{revtex4}
\usepackage[dvips]{graphicx}
\begin{document}

\title{Strongly correlated crystal-field approach to Mott
insulator LaCoO$_{3}$}

\author{R.J. Radwanski}
\homepage{http://www.css-physics.edu.pl}
\email{sfradwan@cyf-kr.edu.pl}
\affiliation{Center for Solid State Physics, S$^{nt}$Filip 5, 31-150 Krakow, Poland,\\
Institute of Physics, Pedagogical University, 30-084 Krakow,
Poland}
\author{Z. Ropka}
\affiliation{Center for Solid State Physics, S$^{nt}$Filip 5,
31-150 Krakow, Poland}

\begin{abstract}
Our success in description of recent electron-spin-resonance
results on Mott insulator LaCoO$_3$, Phys. Rev. B 67 (2003)
172401, lies in taking into account strong electron correlations
among $d$ electrons and the relativistic spin-orbit coupling. In
the developed by us Quantum Atomistic Solid State Theory (QUASST)
we assume that the atomic-like integrity of the 3d$^6$ system is
preserved in the Co$^{3+}$ ion in LaCoO$_3$ and that intra-atomic
correlations are much stronger than crystal field interactions. We
conclude that in LaCoO$_{3}$ there is no intermediate spin state
as came out from band-structure calculations. The excited states
originate from the high-spin $^5T_{2g}$ term, being 12 meV above
the ground $^1A_1$ state. We are convinced that many-electron CEF
approach with strong correlations and the atomic-scale orbital
magnetism is physically adequate approach to 3$d$ oxides.\\
Keywords: Mott insulator, crystal field, spin-orbit coupling,
LaCoO$_3$
\end{abstract}
\maketitle
LaCoO$_{3}$ attracts much attention in recent 50 years
due to its insulating non-magnetic ground state and the
significant violation of the Curie-Weiss law at low temperatures
in the temperature dependence of the paramagnetic susceptibility
$\chi(T)$ exhibiting a pronounced maximum at 100 K \cite{1}. This
$\chi(T)$ dependence is customarily discussed in terms of
successive changes of spin states with the increasing temperature
as low-spin (LS), intermediate-spin (IS) and high-spin (HS) states
\cite{2,3,4,5}.

Recent experimental observation of well-defined localized states
by Noguchi et al. \cite{6} by means of Electron Spin Resonance
(ESR) measurements sharply contradict, according to us,
band-structure calculations \cite{2,3,4,5}, that yield the
continuous energy spectrum for 3$d$ states spread over 6 eV. The
developed by us Quantum Atomistic Solid State Theory \cite{7,8},
incorporating the many-electron crystal-field theory was the only
present theory that was prepared for the ESR results revealing
discrete energy states. We have provided \cite{9} a complete
explanation of Noguchi et al.'s results making use of our earlier
long lasting studies on the CEF-field effect in combination with
the spin-orbit coupling on properties of 3$d$ oxides and of
LaCoO$_{3}$ \cite{10,11,12}.

In the ESR experiment Noguchi et al. \cite{6} have derived $g$
tensor, $g_{\parallel }$=3.35, $g_{\perp }$=3.55, of the excited
quasi-triplet described by an effective spin $\tilde{S}$=1 with a
splitting $D$ = +4.90 cm$^{-1}$ (=7.056 K=0.6 meV) with external
magnetic field applied along different crystallographic directions
of single-crystalline LaCoO$_{3}$. We have proven that these
localized states originate from the $^{5}D$ term and that the $g$
tensor and the behavior of excitation energies in fields up to 60
T is very well described by the crystal-field Hamiltonian with
taking into account the spin-orbit interactions \cite{9}.
\begin{figure}[ht]
\centering
\includegraphics[width = 8 cm]{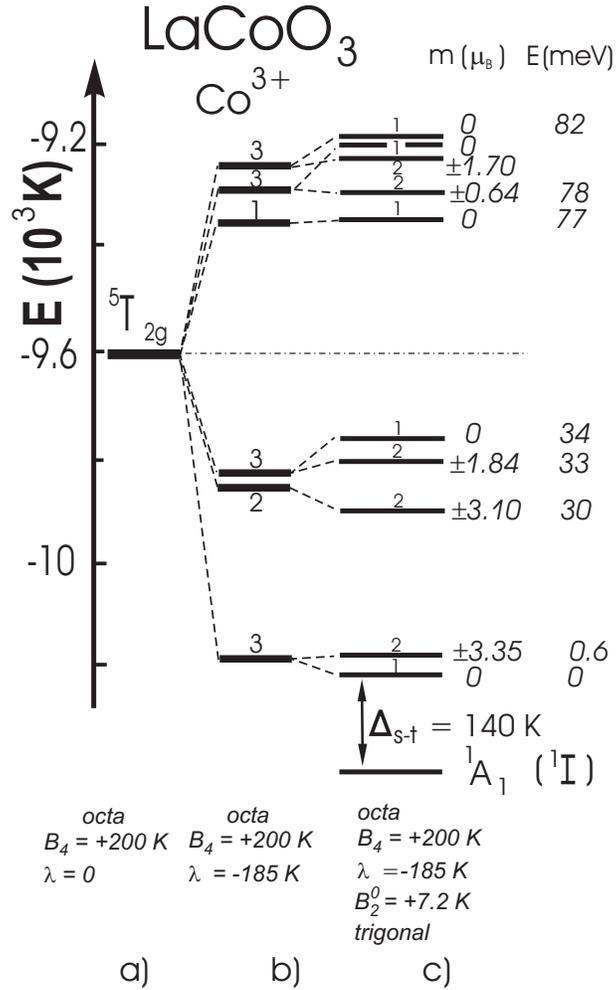}
\caption{Calculated low-energy localized states of the Co$^{3+}$
ion in LaCoO$_{3}$ originating from the $^{5}T_{2g}$ subterm with
the ground state $^{1}A_{1}$.}
\end{figure}

The aim of this paper is to discuss the role of electron
correlations and their manifestation in LaCoO$_{3}$ as well as the
physics of low-spin (LS), intermediate-spin (IS) and high-spin
(HS) states.
\begin{figure}[th]
\centering
\includegraphics[width = 10 cm]{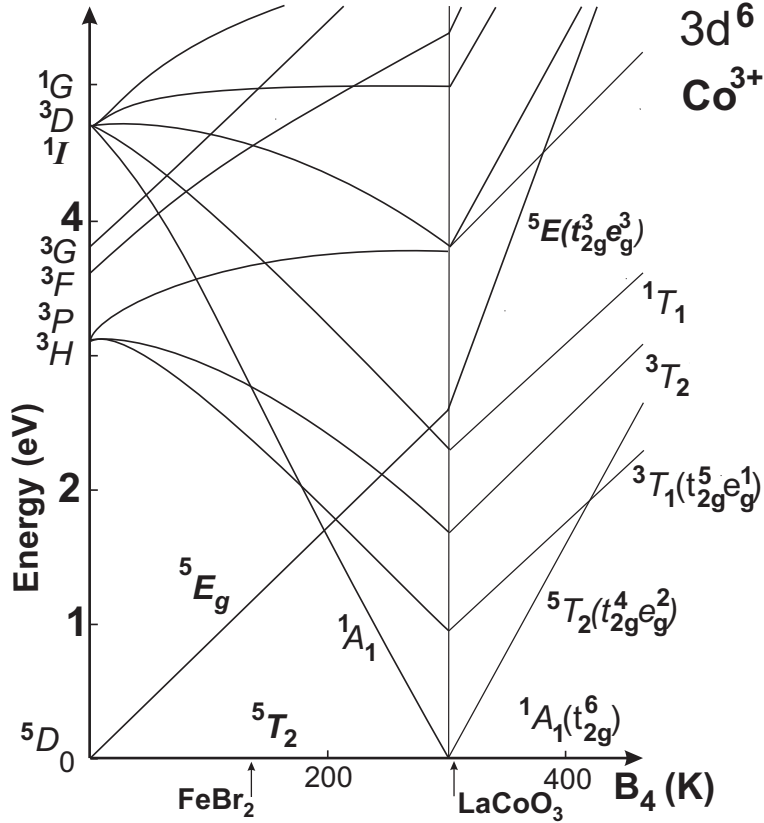}
\caption{Influence of the strength of the octahedral crystal field
on the atomic state of the Co$^{3+}$ ion (3d$^{6}$ system)
(Tanabe-Sugano diagram) \cite{13}. Data for FeBr$_{2}$ are after
Ref. \cite{14}.}
\end{figure}
We have proven that the excited triplet state originates from the
atomic-like high-spin ($S$=2) term $^{5}D$ of the 3$d^{6}$ electron
system existing in the Co$^{3+}$ ion because i) such the triplet is
expected for the octahedral crystal field in the presence of the
spin-orbit interactions \cite{10}, ii) by perfect reproduction of
the $g$ tensor and iii) the splitting of the triplet by the
rhombohedral distortion as well as iv) the reproduction of the
behavior of excitation energies in fields \textbf{B} up to 60 T for
different crystallographic directions with the single-ion
Hamiltonian for the term $^{5}D$ with $S$=2 and $L$=2:
\begin{equation}
H_{d}=H_{cub}(L,L_{z})+\lambda_{s-o} L\cdot
S+B_{2}^{0}O\,_{2}^{0}(L,L_{z})+\mu _{B}(L+g_{s}S)\cdot \textbf{B}
\end{equation}
The good description proves the energy scales assumed in QUASST for
consideration of 3$d$ oxides: the dominant energy is the octahedral
CEF interactions (2-3 eV), weaker the spin-orbit coupling (16 meV)
and the weakest are off-octahedral distortions (1 meV).

In the shown description we use only 3 parameters: B$_{4}$ (from
\emph{ab initio} calculations for the $d^{6}$ configuration
B$_{4}$ turns out to be positive for the O$^{2-}$ octahedron with
the found value of 280-300 K), $\lambda_{s-o} $ and B$_{2}^{0}$.
All of them have clear physical meaning.
\begin{figure}[ht]
\centering
\includegraphics[width = 8 cm]{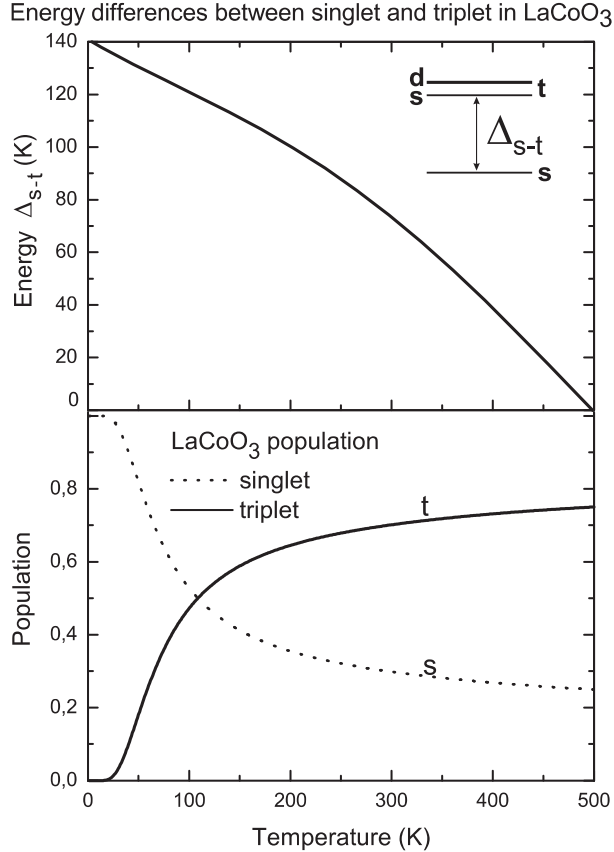}
\caption{Temperature dependence of the energy separation
$\Delta_{s-t}$ between the lowest singlet and the excited triplet.
This dependence is inferred from the thermal expansion and
experimental evaluation of temperature dependence of the Co-O
distance. In the bottom part the calculated results for the
population of the singlet and triplet states is shown. The
population of the triplet state can be understood as population of
the Co$^{3+}$ ions in the high-spin state.}
\end{figure}

Detailed electronic structure of LaCoO$_{3}$ is shown in Fig. 1.
At lowest temperatures the only populated state is $^{1}A_{1}$
subterm that is a singlet. It is the reason for the diamagnetism
of LaCoO$_{3}$ as all atoms are in singlet ground states. The
electronic structure below 25 meV determines magnetic and
electronic properties below room temperatures.

We conclude that:

i) LaCoO$_{3}$ is the Mott insulator. It is insulator, despite of
the incomplete 3$d$ shell, due to strong atomic-like electron
correlations, that in QUASST we recognize as predominantly
intra-atomic.

ii) There is no intermediate spin state in contrast to
band-structure calculations of Ref. 2 and numerous further
quotations.

iii) The low spin state is the singlet $^{1}A_{1}$ subterm - it
originates from the atomic term $^{1}I$ and becomes the ground
state by octahedral crystal-field interactions due to the large
orbital moment ($L$ = 6) of the $^{1}I$ term (Fig. 2).

iv) Detailed description of magnetic and electronic properties of
LaCoO$_{3}$ is hampered as the electronic structure shown in Fig.
1 changes with temperature. Due to the thermal expansion the
energy separation $\Delta_{s-t}$ decreases becoming zero at,
according to our estimations, about 500 K, Fig. 3. The applied
pressure causes an increase of $\Delta_{s-t}$ what will manifest
in the shift of the $\chi(T)$ maximum to higher temperatures. With
increasing temperature the population of the high-spin state
rapidly grows - already at 100 K the LS- and HS-state population
becomes equal. Above 100 K the HS state is more populated than LS
state.

v) So fast population of the HS state enables the development of
the orbital ordering with temperature as the Co ions in the HS
state have non-spherical shape of their charge cloud.

vi) We think that the Co sublattice can lose homogenity and 2/3 of
Co ions experience stronger the octahedral crystal field and have
larger value of $\Delta_{s-t}$ than, say 150 meV. Due to this
larger $\Delta_{s-t}$ they only slightly contribute to the
macroscopically-observed susceptibility. But this point needs
further studies.

\end{document}